\documentclass[aps,prl,amsmath,twocolumn,groupedaddress,showpacs,letterpaper]{revtex4}

\usepackage{graphicx}

\begin{document}
\title{The Single-Photon Router}

\author{Io-Chun Hoi$^{1}$,C.M. Wilson$^{1}$,G\"oran Johansson$^{1}$,Tauno Palomaki$^{1}$,Borja Peropadre$^{2}$ \& Per Delsing$^1$}


\affiliation{$^1$Department of Microtechnology and Nanoscience (MC2), Chalmers University of Technology, SE-412 96 G\"oteborg, Sweden}
\affiliation{$^2$Instituto de F\'isica Fundamental Serrano, CSIC, Madrid, Spain}

\begin{abstract}
We have embedded an artificial atom, a superconducting ``transmon" qubit, in an open transmission line and investigated the strong scattering of incident microwave photons ($\sim6$ GHz). When an input coherent state, with an average photon number $N\ll1$ is on resonance with the artificial atom, we observe extinction of up to 90\% in the forward propagating field. We use two-tone spectroscopy to study scattering from excited states and we observe electromagnetically induced transparency (EIT). We then use EIT to make a single-photon router, where we can control to what output port an incoming signal is delivered. The maximum on-off ratio is around 90\% with a rise and fall time on the order of nanoseconds, consistent with theoretical expectations. The router can easily be extended to have multiple output ports and it can be viewed as a rudimentary quantum node, an important step towards building quantum information networks.
\end{abstract}

\date{\today}
 \pacs{42.50.Gy, 85.25.Cp, 03.67.Hk}          
 \maketitle     
In recent years, quantum information science has advanced rapidly, both at the level of fundamental research and technological development.  For instance, quantum cryptography systems have become commercially available \cite{Scarani}.  These systems are examples of quantum channels, serving mainly to distribute quantum information.  There is a significant effort to combine these quantum channels with quantum nodes that would offer basic processing and routing capability.  The combination of these channels and nodes would create a quantum network enabling applications simply impossible today \cite{Kimble}.  Quantum networks connecting simple quantum processing nodes are also a promising architecture for a scalable quantum computer.  In this letter, we demonstrate a rudimentary quantum node, a single-photon router. The active element of the router is a single "artificial atom", a superconducting qubit, strongly coupled to a superconducting transmission line.  Exploiting the phenomenon of electromagnetically induced transparency (EIT), we show that we can route a single-photon signal from an input port to either of two output ports with an on-off ratio of 90\%. The switching time of the device is shown to be a few nanoseconds, consistent with theoretical expectations and the device parameters.  The device is a nanofabricated circuit offering a clear path to scalability. For instance, it is straight forward to extend this router to select between multiple channels. 

An obvious requirement of a quantum channel is the ability to coherently distribute quantum information over relatively large distances.  This typically implies the use of photons as information carriers, as opposed to other quantum systems such as atoms.  This presents a problem when trying to implement a quantum node, as the interaction of photons with themselves is vanishingly small.  Without interactions, photons cannot be controlled or directed.  We can however look to control the photons by using matter as an intermediary, exploiting the strong interactions of electrons for instance.  Still, the coupling of single-photon signals to bulk nonlinear materials is too weak for efficient control.  A number of authors \cite{Shen1,Shen2,Zumofen,Witthaut} have suggested that this problem could be overcome by resonantly coupling the signals to single atoms, which are highly nonlinear systems.  While impressive technical achievements have been demonstrated in experiment, the coupling of single atoms to light remains relatively inefficient.  For instance, in the prototypical experiment of scattering light from a single atom, the reduction in the intensity (extinction) of the incident light does not exceed $~12\%$  \cite{Tey,Hwang,Wrigge,Gerhardt}.  However, it was recently demonstrated that microwave photons can be coupled extremely efficiently to a single artificial atom, showing extinction efficiencies in an open transmission line greater than $~90\%$ \cite{Astafiev1,Abdumalikov}.  This approach is the basis of our single-photon router.

Our artificial atom is a superconducting transmon \cite{Koch} qubit, consisting of two Josephson junctions in a SQUID configuration with a total capacitance $C_{\Sigma}$. It is capacitively coupled to a 1D transmission line (see Fig. 1(a)) in a coplanar waveguide configuration. The two lowest energy states $\left\vert 0 \right\rangle$,$\left\vert 1 \right\rangle$ have a transition energy $\hbar\omega_{01}(\Phi)\approx \sqrt{8E_JE_c}-E_c \sim 7.1$ GHz where $E_J$ is the Josephson energy of the SQUID and $E_C$ is the charging energy (see Supplementary for details). This type of qubit has been extensively studied \cite{Schuster,Fink:2008,Sandberg,Mallet:2009} and successfully used to, for example, perform quantum algorithms \cite{Dicarlo} as well as produce single photons\cite{Houck}.   

The electromagnetic field in the transmission line can be described by incoming $(+)$ and outgoing $(-)$ voltage waves on the left $(L)$ and the right $(R)$, $V_{L/R}^{\pm}$. In Fig.1(a), the transmission coefficient $t=V_R^-/V_L^+$  and the reflection coefficient $r=V_L^-/V_L^+$ are related by the definition, $t=r+1$. When the applied probe frequency $\omega_p$ is equal to $\omega_{01}$, the reflection coefficient is given by \cite{Astafiev1}
\begin{eqnarray}
r=-r_0\frac{1}{1+\Omega_p^2/\Gamma_{10}\gamma_{10}},
\end{eqnarray}
where the maximum reflection amplitude is $r_0=1/(1+2\Gamma_{\phi}/\Gamma_{10})$. $\Gamma_{10}$  is the relaxation rate of the ÒatomÓ from $\left\vert 1 \right\rangle$ to $\left\vert 0 \right\rangle$, $\gamma_{10}=\Gamma_{10}/2+\Gamma_{\phi}$ is the 0-1 decoherence rate, and $\Gamma_{\phi}$ is the 0-1 pure dephasing rate. A coherent input signal (probe) will drive coherent oscillations of the atom at a Rabi frequency which is linear in the probe amplitude and can be written as \cite{Koch}
\begin{eqnarray}
\Omega_p=\frac{2e}{\hbar}\frac{C_c}{C_{\Sigma}}\left(\frac{E_J}{8E_c}\right)^{1/4}\sqrt{PZ_0}
\end{eqnarray}
where $P=|V_L^+|^2/2Z_0$ is the probe power, $Z_0\sim50$ $\Omega$  is the impedance of the transmission line and $C_c$ is the coupling capacitance between the transmon qubit and the transmission line (Fig. 1(b)). For a weak resonant probe $(\Omega_p<<\Gamma_{10},\gamma_{10})$ and in the absence of both pure dephasing $(\Gamma_{\phi}=0)$ and unknown loss channels, such as relaxation of the qubit not associated with the coupling to the transmission line, we expect to see full reflection of the incoming probe\cite{Chang,Zumofen,Shen2}. This perfect reflection can be understood as the coherent interference between the incoming wave and the wave scattered from the atom.

\begin{figure}
 \includegraphics[width=\columnwidth]{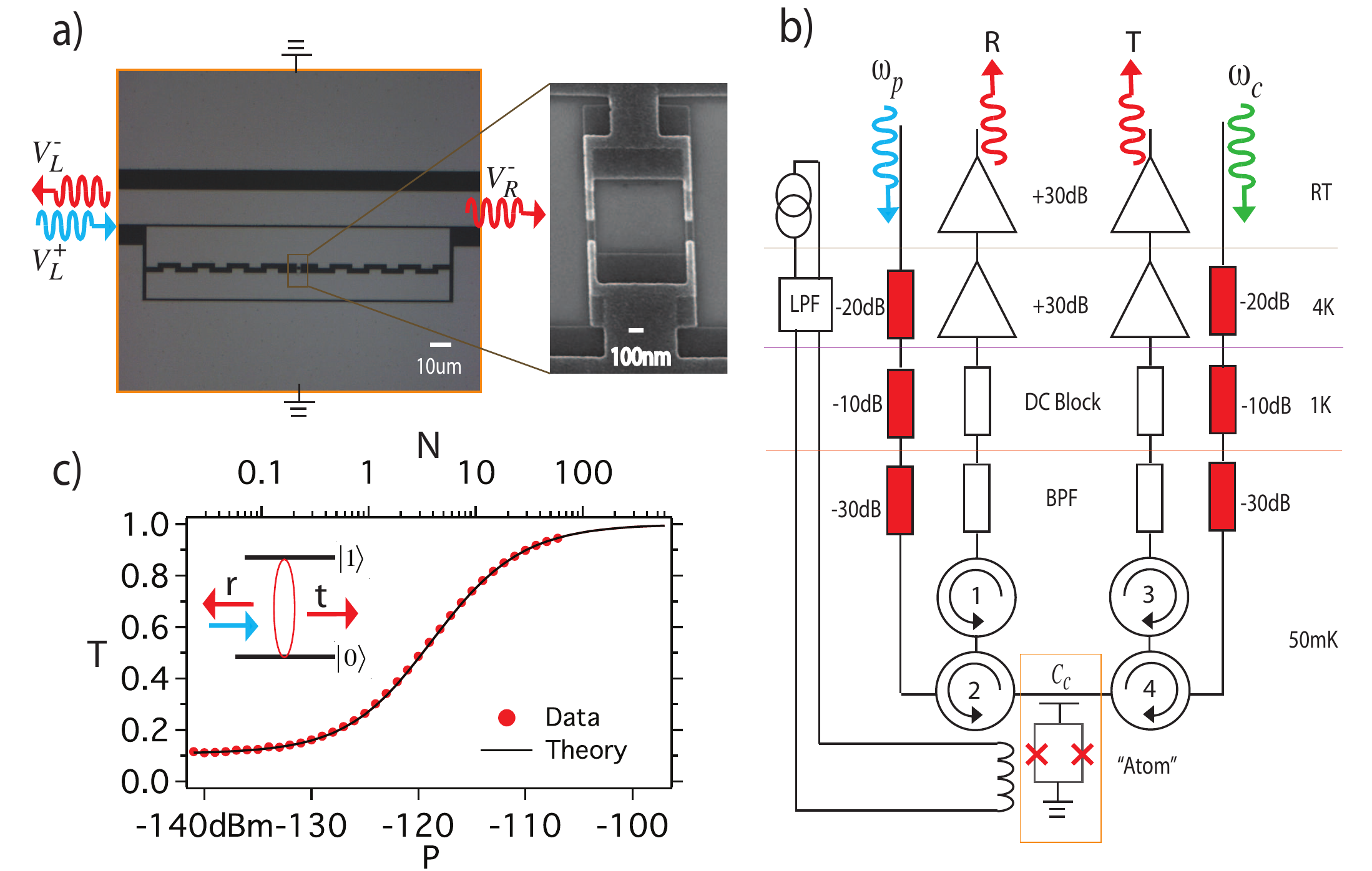}
\caption{Scattering from a single artificial atom. (a) A  micrograph of our artificial atom, a superconducting transmon qubit embedded in 1D open transmission line. (Zoom In) Scanning-electron micrograph of the SQUID loop of the transmon, which allows us to tune the transition frequency of the transmon with an external magnetic flux $\Phi$. (b) Schematic of the measurement setup. A strong control pulse at $\omega_{c}=\omega_{12}$  is used to route a weak microwave signal at the probe frequency $\omega_{01}$.  We measure the transmitted and reflected probe simultaneously in the time domain. (c) Transmittance $T=|t|^2$ as a function of incident power. On the top axis is the average photon number $N\equiv P/(\hbar\omega_p(\Gamma_{10}/2\pi))$, with $-124$ dBm corresponding to one photon per interaction time, $2\pi/\Gamma_{10}$. 90$\%$ of the coherent photons are not transmitted in the single photon regime, defined as $N<1$. (Inset) A weak, resonant coherent state is scattered by the atom.}
\end{figure}

We have measured the reflection and transmission coefficients as a function of incident probe power using homodyne detection (See Supplementary). By fitting the frequency and power dependence of this data we extract $E_J/h = 12.7$ GHz, $E_c/h = 590$ MHz, $\Gamma_{10}/2\pi=73$ MHz and $\Gamma_{\phi}/2\pi=18$ MHz.  In Fig. 1(c), we plot the transmittance $T=|t|^2$ on resonance as a function of $P$. We see an extinction of propagating photons of up to 90\% for photon numbers $N\ll1$. The strong saturation of the extinction already at single-photon powers is an indication that the scattering is caused by a single atom, since the atom can only absorb and emit one photon at a time. While we show only the transmitted data for these graphs, the reflected power also varies with power as expected.

\begin{figure*}
 \includegraphics[width = 2\columnwidth]{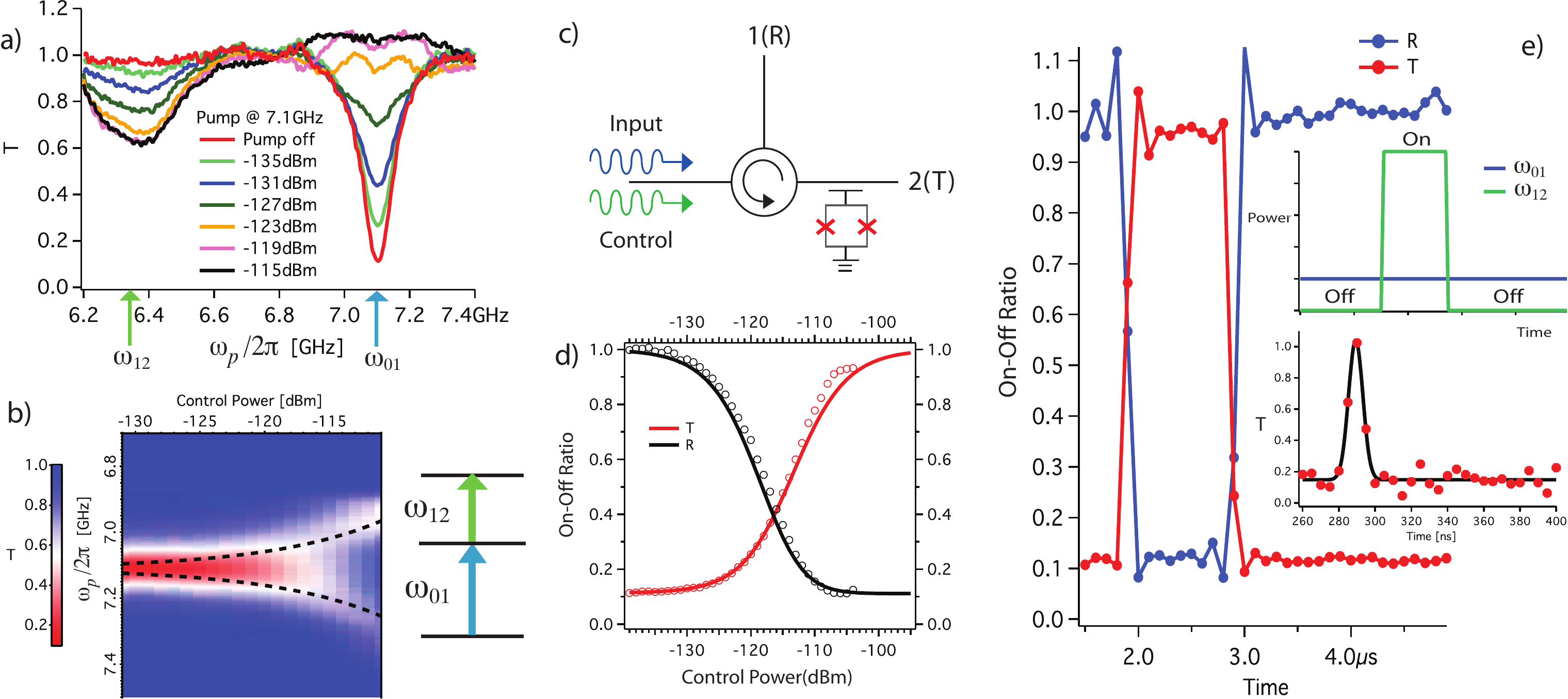}
\caption{(a) A microwave pump is continuously applied at  $\omega_{01}$ with increasing power while a weak probe tone is swept in frequency. As the population in the first state is increased, due to the drive at $\omega_{01}$, scattering at $\omega_{12}$  becomes possible, appearing as another dip in the transmittance. From this, we extract $\omega_{12}/2\pi=6.38$ GHz.  (b) The microwave pump is now applied at $\omega_{12}$.  As the power of the $\omega_{12}$ pump increases, we see electromagnetically induced transparency (EIT) at $\omega_{01}$. At strong drive, the Aulter-Townes doublet appears with a splitting equal to the Rabi frequency $\Omega_{c}/2\pi $ (black dashed lines).  (inset) Energy level diagram. (c) Cartoon of the router. With the control off, the input probe is reflected from the transmon, and is routed to port 1 through the circulator.  When the control is on, the input is transmitted to port 2.   (d) Normalized on-off ratio (see text) of the transmittance and reflectance as a function of control pulse power, measured simultaneously. The circles are the data and the solid lines are fits using Eqs. 3-5. (e) Time dependence of the transmittance and reflectance at $\omega_{01}$, measured simultaneously, while a control pulse is applied. We see that the input signal is routed with an on-off ratio of $\sim90$\%.  The upper inset shows the control pulse sequence. The lower inset shows the response to a 10 ns Gaussian control pulse (circles), along with a Gaussian fit (solid line). We see that the transmittance smoothly follows the control on the few ns timescale while maintaining the high on-off ratio.}
\end{figure*}

So far, we have considered only the lowest two energy levels of our artificial atom. In reality, the transmon has several higher states, in particular it has a second excited state with the 1-2 transition frequency $\omega_{12}$ . This second transition can be directly measured using 2-tone spectroscopy (See Fig. 2(a)). We extract $\omega_{12}/2\pi=6.38$ GHz, giving an anharmonicity of $\alpha = 720$ MHz between the two transitions.  The linewidth of the 1-2 transition is dominated by the charge dispersion of $\left\vert 2 \right\rangle$. Further increasing the pump power, we observe the well known Mollow triplet \cite{Astafiev1,Baur,Mollow} (data not shown).  The Rabi splitting of the triplet was then used to calibrate the applied microwave power at the ÒatomÓ.
 
By pumping the system at  $\omega_{c}=\omega_{12}$, we can observe the phenomenon of electromagnetically induced transparency (EIT)\cite{Fleischhauer,Abdumalikov,Kelly,Slodicka}.  With the pump off, we have seen that an incident, low-power probe at $\omega_{01}$ is reflected.  When the pump is applied, quantum interference suppresses the 0-1 transition. Therefore, the atom becomes transparent to the probe beam at $\omega_{01}$.  We show this in Fig. 2(b), where the probe transmission is plotted as a function of probe frequency and control power.  We see in fact that for large control powers, the original line splits into a doublet with a separation of $\Omega_c$.  This is known as the Aulter-Townes doublet \cite{Autler}.
 
We exploit EIT to create a single-photon router. The operation principle is explained as follows (see Fig. 2). We input a weak, continuous probe in the single-photon regime at $\omega_{01}$. We then apply a strong control pulse, around 30dB stronger than the probe, at  $\omega_c = \omega_{12}$.  When the control is off, the photons are reflected by the atom and travel through the circulator to output 1. When the control is on, the photons are transmitted due to EIT, and travel to output 2. The setup is shown in Fig. 1(b), which enables us to measure the reflected and transmitted probe power simultaneously in
the time domain. This is crucial to demonstrate that the extinction of the transmitted beam is due to reflection instead of loss. One could also envision making a photon router by simply detuning the 0-1 transition of the atom via magnetic flux through the SQUID loop. However, the power needed to generate sufficient flux to detune our atom is several orders of magnitude higher \cite{Sandberg}. 

The operation of the router is demonstrated in Fig. 2(e). As expected, when the control is on, most of the signal is transmitted while little is reflected. A nearly 90$\%$ on-off ratio (defined below) is achieved in both the reflectance and transmittance of the coherent signal.  In the setup of Fig.~1(b), we send down $\omega_{p}$ and $\omega_{c}$  in opposite directions with respect to the artificial atom. We see very similar results if they are instead input through the same port. 
                       
 We also characterized the time response of the router. For a control pulse width greater than or equal to 50 ns, we used a square pulse.  For shorter control pulses, we use a gaussian pulse. We see a constant on-off ratio of approximately 90$\%$ down to the shortest pulses, which had a full width at half maximum of 10 ns. The time resolution of our microwave digitizer is 5ns, preventing us from accurately measuring pulses shorter than this. Still, we see that the transmission follows the control on the few ns time scale, limited by our instrumentation.  We would expect the limit of the device to be $\sim1/\Gamma_{01}=2$ ns.

                                              
  In Fig. 2(d), we characterize the on-off ratio as a function of control power. For a probe power in the single-photon regime with the control and probe on resonance, the transmission of the probe, for a control amplitude corresponding to $\Omega_c$, is \cite{Abdumalikov}
\begin{eqnarray}
t=1-\frac{\Gamma_{10}}{2\gamma_{10}+\frac{\Omega_c^2}{2\gamma_{20}}},
\end{eqnarray}
where $\gamma_{20}$ is the dephasing rate of the 0-2 transition. The dipole moment of the 1-2 transition is $\sqrt{2}$ larger than that of the 0-1 transition \cite{Koch}, so for a particular microwave power, we have $\Omega_c(P)=\sqrt{2}\Omega_p(P)$. We define the coherent transmittance and reflectance as $T(\Omega_c)=|t|^2$ and $R(\Omega_c)=|1-t|^2$ , respectively. The missing power, $1-R-T$, comes from  incoherent emission which averages out in amplitude measurements.  We can then define the normalized on-off ratios
\begin{eqnarray}
R_{on\_off}=\frac{R(\Omega_c)+R_b}{R(0)+R_b};
\frac{T_{on\_off}}{T_{o}}=T(\Omega_c).
\end{eqnarray}
where $R_{b}$ accounts for background reflections in the line and leakage through circulator 2 (Fig 1(b)) and $T_{o}$ is the transmission measured far detuned from the transmon transitions.  By using the same value of $\Gamma_{10}$ and $\gamma_{10}$  as before, together with the new fitting parameter $\gamma_{20}/2\pi=145$ MHz  and $R_b=0.05$ , we get the solid lines in Fig. 2(d). These values agree with our expectations based on the charge dispersion of $\left\vert 2 \right\rangle$ and circulator leakage in our system.

The performance of our single-photon router can be improved in both operation speed and routing efficiency. An increased coupling, $\Gamma_{10}$, between the transmon and the transmission line would improve the operation speed.  This can be done by simply increasing the coupling capacitance, $C_c$.  The ultimate limit of $\Gamma_{10}$ is the anharmonicity $\alpha$, which was 720 MHz for this sample, typical of transmon qubits.  This implies switching times below 1 ns should be feasible.  Within the current model, the switching efficiency is limited by the maximum reflectance $R_{max} = (\Gamma_{10}/2\gamma_{10})^2\approx1-4\Gamma_{\varphi}/\Gamma_{10}$, which is limited by pure dephasing.  Optimized transmons have been demonstrated to have $\Gamma_{\varphi}/2\pi\sim30$ kHz \cite{Schreier}.  With this dephasing rate and a modestly enhanced $\Gamma_{10}/2\pi = 100$ MHz, we would get an efficiency of 99.9\%.  These low dephasing rates where however measured for transmons in cavities, which serves to protect the transmon from environmental noise.  We might expect higher dephasing in a completely open transmission line.  Since dephasing is caused by low-frequency noise, a high-pass filter in the line would serve to protect the transmon in the same way without effecting the response at the qubit frequency.

\begin{figure}
 \includegraphics[width=\columnwidth]{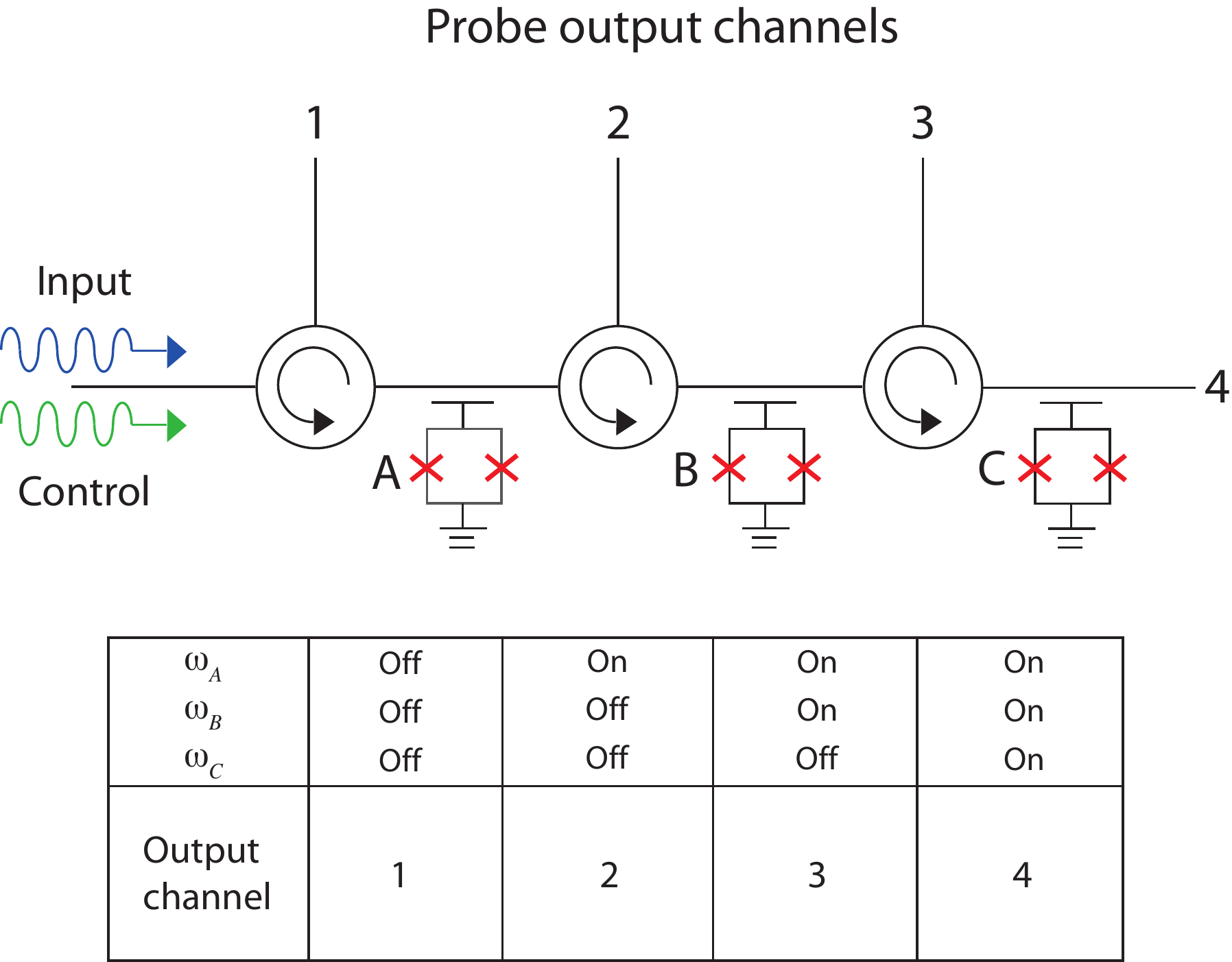}
\caption{Top: Cartoon of a multiport router. The router demonstrated here can easily be cascaded to distribute photons to many channels.  Here we show a 4 port router using 3 atoms (A,B,C) in series, each separated by a circulator. The 0-1 transition frequencies of the atoms are the same, while the 1-2 transition frequencies, $\omega_A \ne \omega_B \ne \omega_C$, are different. This arrangement can be designed in a straightforward manner by controlling the ratio of $E_J/E_c$.  By turning on and off control tones at the various 1-2 transition frequencies, we can determine the output channel of the probe field, according to the table. For instance, if we wish to send the probe field to channel 3, we apply two control tones at $\omega_{A}$ and $\omega_{B}$. We note that all the control tones can be input through the same port regardless of the number of output channels, reducing the complexity of the design. Bottom: Control table to select a given output port. }
\end{figure}

In conclusion, we have demonstrated a basic single-photon router with high speed and efficiency.  The device can be viewed as a rudimentary quantum node.  As shown in Fig. 3, the operation scheme used for the device is scalable in a straightforward manner and could see use in future quantum networks.

We acknowledge financial support from the Swedish Research Council, the Wallenberg foundation and from the European Research Council. We would also like to acknowledge O. Astafiev for fruitful discussions and  M. Sandberg and F. Persson for fabrication and experimental help.


\end{document}